\documentclass[aps,prb,superscriptaddress,reprint,floatfix]{revtex4-2}
\usepackage{hyperref}
\usepackage[english]{babel}
\usepackage{amsmath,amssymb,upgreek}
\usepackage[separate-uncertainty = true,multi-part-units=single]{siunitx}
\usepackage{color,graphicx}
\usepackage{soul,enumitem}
\usepackage{multirow}

\usepackage[calcwidth,explicit]{titlesec}
\titleformat{\section}{\bfseries\sffamily\large\filcenter}{\thesection.}{0.3em}{\MakeUppercase{#1}}
\titlespacing{\section}{0pt}{1ex}{0.5ex}
\titleformat{\subsection}{\bfseries\sffamily\filcenter}{\thesubsection.}{0.2em}{#1}
\titlespacing{\subsection}{0pt}{0.8ex}{0.2ex}
\titlespacing*{\paragraph}{0em}{0ex}{0.35em}[]
\titleformat{\paragraph}[runin]{\normalfont\normalsize\bfseries}{}{0pt}{}
\renewcommand\thesection{\Alph{section}}%\renewcommand\thesubsection{\Alph{section}\Arabic{subsection}}

\makeatletter\@addtoreset{paragraph}{section}\makeatother
\makeatletter\def\p@paragraph{}\makeatother

\begin{document}%
\title{Quantifying symmetric exchange in ultrathin ferromagnetic films with chirality} %
%
%\date{July 2022}%
%
%
\author{Tobias Böttcher}%
\affiliation{Fachbereich Physik and Landesforschungszentrum OPTIMAS, Technische Universit\"at Kaiserslautern, 67663 Kaiserslautern, Germany}%
\affiliation{MAINZ Graduate School of Excellence, 55128 Mainz, Germany}%
\author{T. S. Suraj}\thanks{These authors contributed equally to this work}
\affiliation{Department of Physics, National University of Singapore, 117551, Singapore}
\author{Xiaoye Chen}\thanks{These authors contributed equally to this work}%
\affiliation{Institute of Materials Research \& Engineering, Agency for Science, Technology \& Research, 138634, Singapore}%

\author{Banibrato Sinha}%
\affiliation{Fachbereich Physik and Landesforschungszentrum OPTIMAS, Technische Universit\"at Kaiserslautern, 67663 Kaiserslautern, Germany}%
\altaffiliation{Current affiliation: Department of Applied Physics, Northwestern University, Evanston, IL 60208, USA}%
\author{Hui Ru Tan}
\affiliation{Institute of Materials Research \& Engineering, Agency for Science, Technology \& Research, 138634, Singapore}%
\author{Hang Khume Tan}%
\affiliation{Institute of Materials Research \& Engineering, Agency for Science, Technology \& Research, 138634, Singapore}%
\author{Robert Laskowski}%
\affiliation{Institute of High Performance Computing, Agency for Science, Technology \& Research, 138632, Singapore}
\author{Burkard Hillebrands}%
\affiliation{Fachbereich Physik and Landesforschungszentrum OPTIMAS, Technische Universit\"at Kaiserslautern, 67663 Kaiserslautern, Germany}%
\author{Mikhail Kostylev}%
\affiliation{School of Physics, M013, University of Western Australia, 6009 Perth, Australia}%
\author{Khoong Hong Khoo}%
\affiliation{Institute of High Performance Computing, Agency for Science, Technology \& Research, 138632, Singapore}

\author{Anjan Soumyanarayanan}%
\altaffiliation{Corresponding authors:\\ anjan@imre.a-star.edu.sg, ppirro@physik.uni-kl.de}%
\affiliation{Institute of Materials Research \& Engineering, Agency for Science, Technology \& Research, 138634, Singapore}%
\affiliation{Department of Physics, National University of Singapore, 117551, Singapore}
\author{Philipp Pirro}%
\altaffiliation{Corresponding authors:\\ anjan@imre.a-star.edu.sg, ppirro@physik.uni-kl.de}%
\affiliation{Fachbereich Physik and Landesforschungszentrum OPTIMAS, Technische Universit\"at Kaiserslautern, 67663 Kaiserslautern, Germany}%
\begin{abstract} 
The symmetric (Heisenberg) exchange interaction is fundamental to magnetism and assumes critical importance in designing magnetic materials for novel emergent phenomena and device applications. However, quantifying exchange is extremely challenging for ultrathin ($\sim$ 1~nm) magnetic films, as techniques and approximations reliably used for bulk materials are largely inapplicable in the two-dimensional (2D) limit.
Here we present and contrast the measurement of exchange stiffness, $A$, by several methods on a series of five Co/Pt-based ultrathin ($1-2$~nm) films. We compare results from (a) spin-wave spectroscopy by Brillouin light scattering (BLS), (b) three analytical models describing the temperature dependence of magnetization obtained by magnetometry, (c) microscopic domain periodicity measurements and simulations, and (d) ab initio density functional theory (DFT) calculations.
While different methods present some qualitatively consistent trends across samples, we note, for any given sample, considerable differences (up to $5\times$) in the absolute values of $A$ across the techniques, consistent with discrepancies of $A$ reported in literature for nominally similar samples. 
We analyze possible sources of the discrepancies across various methods, notably including their relationship to the spin-wave dispersion, and the wave-vector ranges probed. We compare the strengths and limitations of the techniques, and outline directions for their future use in characterizing exchange interactions in ultrathin films.
\end{abstract}%
\maketitle%
\section{Introduction}\label{Sec_Intro}
%%%% Introduction 

% A1: Motivation for exchange interaction in ultrathin films
\paragraph{Motivation}
In recent years, there has been rapid growth of interest on understanding the behavior of magnetic films with thicknesses approaching the two-dimensional (2D) or atomic limit. On one hand, multilayered films interfacing such ultrathin ferromagnets (FMs) with heavy metals give rise to new, emergent phenomena – including chirality, topology, and spin-charge conversion \cite{Hellman.2017,Manchon.2015,Soumyanarayanan.2016}. On the other hand, spintronic devices developed from such ultrathin multilayers have attractive properties for memory and computing applications \cite{Parkin.2008,Fert.2017,Hirohata.2020,Dieny.2020}. 

% A2: Magnetic Interactions
\paragraph{Magnetic Interactions}
Emergent phenomena in ultrathin magnetic films are governed by magnetic interactions arising at the atomic scale \cite{Hellman.2017}. First, the exchange stiffness, $A$, which, within the micromagnetic description, represents the direct (Heisenberg) exchange interaction between neighbouring spins, and characterizes the overall strength of FM order \cite{Bland.2005,Vaz.2008}. Second, the effective anisotropy, $K_{\rm eff}$, which determines the energetically favoured FM orientation – in-plane (IP), or out-of-plane (OP) – and includes crystalline, shape, and interfacial contributions \cite{Johnson.1996,Vaz.2008}. Third, the Dzyaloshinskii-Moriya interaction (DMI), $D$, which arises from interfacial effects in asymmetric multilayers, and endows ultrathin magnets with chirality\cite{Dzyaloshinskii.1958,Moriya.1960,Kuepferling.2020}. The interplay of these interactions may result in a ground state comprising non-collinear nanoscale spin textures, including chiral domain walls and skyrmions\cite{Nagaosa.2013,Jiang.2015,MoreauLuchaire.2016,Woo.2016,Boulle.2016,Soumyanarayanan.2017}. At larger lengthscales ($\gtrsim 100$~nm), this may additionally include contributions from magnetostatic, or dipolar effects\cite{Fert.2017}. In order to design material systems with desired ground states or functional devices with specified characteristics, it is imperative to quantify these key interactions within ultrathin magnetic films.  

%% A3: Measuring Exchange & Limitations for Ultrathin Materials
\paragraph{Current Work \& Limitations}
The anisotropy, $K_{\rm eff}$, of ultrathin films can be measured straightforwardly via magnetometry or microwave spectroscopy \cite{Johnson.1996}, while the DMI, $D$ is typically determined via the asymmetry in spin-wave or domain propagation (some of which implicitly require $A$) \cite{Kuepferling.2020}. In contrast, despite being the most fundamental of the interactions, the exchange stiffness, $A$ is extremely challenging to quantify in the ultrathin limit for several reasons. First, conventional methods used to determine $A$ for thick films -- e.g., via resonance modes  \cite{Schreiber.1996,Klinger.2015} -- are either inapplicable in the 2D limit or cannot be implemented due to signal-to-noise constraints. Second, while magnetometry-based approaches utilizing the Bloch law are commonly used for quantifying $A$ \cite{Shahbazi.2019}, the validity of the underlying three-dimensional (3D) model is questionable for the ultrathin limit, and consensus on mitigating this issue is lacking \cite{Nembach.2015,Yastremsky.2019}.Finally, spin-wave spectroscopy -- widely regarded as the most reliable method \cite{Sebastian.2015} -- has seen limited use for quantifying $A$ in ultrathin chiral films, likely due to constraints imposed by anisotropy, or by dipolar interactions for multiple stack repetitions \cite{Di.2015}. Together, these challenges have resulted in large discrepancies in published values of $A$ for nominally similar ultrathin FMs \cite{Metaxas.2007,Vaz.2008,Yastremsky.2019,Shahbazi.2019}. This lack of consensus is especially concerning given the critical role of $A$ in stabilizing conventional (FM) and novel (chiral) states \cite{Soumyanarayanan.2016,Fert.2017}, governing magnetoresistive memory and spin-wave device characteristics \cite{Dieny.2020,Pirro.2021}, and determining other critical design parameters for functional materials. Therefore, it is imperative to quantitatively benchmark the determination of $A$ values for ultrathin magnetic films across key characterization techniques. 

%% A4: Summary of Results
\paragraph{Results Summary}
In this work, we present and contrast the exchange stiffness $A$ determined for five Co/Pt-based ultrathin films using various methods. First, Brillouin light scattering (BLS) spectroscopy was used to measure the spin-wave dispersion, whose analysis, with supporting measurements, enabled the extraction of $A$.  Next, magnetometry measurements were used to obtain the temperature dependence of magnetization, which were analyzed using different variants of the Bloch model adapted to ultrathin films, which provided an independent estimate of $A$. Furthermore, microscopic domain periodicity measurements, as well as \emph{ab initio} density functional theory (DFT) calculations were also performed for selected cases and compared with experimental results. While qualitative trends across samples are apparent, for a given sample, we note large differences in the magnitude of $A$ obtained by the various methods. We discuss possible origins of these discrepancies, as well as the strengths and limitations of the individual techniques. Finally, we outline principled experimental approaches for the future use of these techniques in characterizing ultrathin films.

\paragraph{Manuscript Structure}
The remainder of this manuscript is structured as follows. In \S\ref{Sec_Methods}, we briefly describe the physical basis for the various characterization methods used to determine $A$ as relevant to ultrathin films; in \S\ref{Sec_Results}, we present the results of these methods applied to a set of five Co/Pt-based ultrathin film samples. Following this, we compare in \S\ref{Sec_Discussion} the results from different methods, and discuss sources of discrepancies, and conclude in \S\ref{Sec_Conclusion} by outlining directions for future quantitative efforts.
\section{Methods for Exchange Determination}\label{Sec_Methods}%
\subsection*{Spin-Wave Dispersion from BLS Spectroscopy}
\paragraph{BLS Introduction}
BLS spectroscopy utilises the inelastic scattering of photons with magnons and is an established method for the experimental investigation of spin wave dispersion in magnetic thin films for wave vectors $k$ up to $\sim$\SI{25}{rad \per \upmu m} \cite{Sebastian.2015}. 
In recent years, BLS spectroscopy has been used extensively to measure the interfacial DMI in chiral multilayers \cite{Belmeguenai.2015,Di.2015,Nembach.2015,Shahbazi.2019}. In contrast, the corresponding symmetric exchange interaction has received scant attention for ultrathin films likely due to the concomitant presence of interlayer interactions which are notably absent in this work.

\paragraph{BLS Model}
The dispersion of spin waves propagating perpendicular to the static magnetization within the plane of a homogeneous, ultrathin film, saturated IP, can be treated within the description by \textcite{Kalinikos.1986}. For chiral magnetic films, interfacial DMI leads to an asymmetry in the spin-wave dispersion, that is linear in $k$ \cite{Belmeguenai.2015}. Meanwhile, the symmetric part of the dispersion allows for the extraction of the Heisenberg exchange stiffness \cite{Bottcher.2021}, and is described by \cite{Di.2015}
\begin{align}
	\label{Eq:Dispersion_sym}
	f(k)_\mathrm{sym} &= \left( {f_\mathrm{SW}(k)+f_\mathrm{SW}(-k)} \right)/{2}  \\
	\nonumber
	&= \frac{\gamma \mu_0}{2 \pi} \Big[ \Big(  H_\mathrm{ext} + \lambda_\mathrm{ex}k^2 + M_\mathrm{S} \mathrm{g}(kt)	\Big) \\
	\nonumber
	&\cdot \Big( H_\mathrm{ext} - H_\mathrm{U} + \lambda_\mathrm{ex}k^2 + M_\mathrm{S} \big(1 -  \mathrm{g}(kt) \big) \Big)  \Big]^{1/2}\,,
\end{align}
where $\gamma$ is the gyromagnetic ratio of the material (see Tbl.~\ref{tab:summary} for values), $t$ is the film thickness and $M_\mathrm{S}$ is the saturation magnetization. The uniaxial anisotropy field $\upmu_0 H_\mathrm{U}= {2K_\mathrm{U}/M_\mathrm{S}}$ is related to the uniaxial anisotropy constant $K_\mathrm{U}$. This should not be confused with the effective anisotropy constant $K_\mathrm{eff}=K_\mathrm{U}-M_\mathrm{S}$ and the effective anisotropy field $H_\mathrm{K}=H_\mathrm{U}-M_\mathrm{S}$.

\paragraph{Exchange from BLS}
The influence of the symmetric exchange is contained in the spin-wave stiffness $\lambda_\mathrm{ex}=2A / (\upmu_0 M_\mathrm{S})$ with the Heisenberg exchange stiffness, $A$. In addition, the term $\mathrm{g}(x) = 1- \left[ 1- \exp (-|x|) \right] / |x|$ accounts for the dipolar interaction between magnetic moments. In order to correctly determine $A$, one needs to duly consider the additional contributions, e.g., from dipolar interactions, anisotropy etc. In particular, the method requires accurate knowledge of $M_\mathrm{S}$ and $\gamma$, which need to be measured using additional methods. Finally, for ultrathin magnetic bilayers (e.g. Fe/Co), the dispersion relation needs to be described using an effective approach for $A$ (see Appendix).

\subsection*{$T$-Dependence of Saturation Magnetization}

\paragraph{Bloch Law for 3D}
The temperature dependence of $M_\mathrm{S}$ of a material, accessible to magnetometry techniques, can be utilized to extract $A$ \cite{Bland.2005,Erickson.1991,Vaz.2008}. Given its extensive usage in this regard, especially for chiral magnetic films \cite{Nembach.2015,Shahbazi.2019}, we examine this technique in detail, including an overview of possible analytical approaches. For temperatures $T$ well below the Curie temperature $T_{\rm c}$, the variation of $M_\mathrm{S}(T)$ results primarily from the excitation of thermal spin waves. This can be understood by analyzing the spin-wave dispersion across frequencies up to the \SI{}{THz} range, with the thermal population of the states described by Bose-Einstein statistics. Within this framework, the dispersion model used and the associated spin-wave density of states (DOS) $\rho(\omega)$ play important roles, where $\omega$ is the spin-wave frequency. For the case of a three-dimensional (3D) “bulk” sample with parabolic dispersion, i.e., $\omega(k) = \gamma \mu_0 \lambda_\mathrm{ex} \cdot k^2=2 \gamma (A/M_\mathrm{S}) \cdot k^2$ (i.e., $\rho(\omega)\propto \sqrt{\omega}$), neglecting dipolar interactions, anisotropies, and DMI, we obtain the well-known Bloch $T^{3/2}$ law \cite{Bloch.1930,Yastremsky.2019}

\begin{equation}
	\label{m_reduction_bloch}
	M(T)_{\mathrm{S},{T^{3/2}}} = M_\mathrm{S}(0) - 2 \upmu_B \cdot \eta \cdot \left( \frac{k_\mathrm{B} T M_\mathrm{S}(0)}{2 \gamma A \hbar} \right)^{3/2}\,,
\end{equation}
where $M_\mathrm{S}(0) \equiv M_\mathrm{S}(T=0)$ and $\eta$ is a dimensionless prefactor.

\paragraph{Bloch Law Limitation for 2D}
The Bloch law is particularly suitable for describing the $M_{\rm s}(T)$ character of bulk samples at low temperatures \cite{Maeda.1973}, where $\eta \simeq 0.0587$ \cite{Vaz.2008,Yastremsky.2019}. In this case, using the Bloch law to determine $A$, via a fit to Eqn.~\eqref{m_reduction_bloch}, requires the quantitative determination of $M_\mathrm{S}(0)$. While Bloch’s $T^{3/2}$ law has been used to quantify exchange for nanometer-thick films \cite{ Shahbazi.2019}, some have questioned its validity in the ultrathin regime, as the reduced dimensionality strongly influences the spin-wave DOS \cite{Seavey.1958,Erickson.1991,Nembach.2015}. Consequently, several works  employed a modified Bloch law \cite{Kipferl.2004,Cojocaru.2014,Nembach.2015}, for instance, by renormalizing the factor $\eta$ to account for the altered magnon density in the ultrathin limit \cite{Nembach.2015}. In the following section, we estimate $A$ for our $t = 1 (2)$~nm films using values determined for NiFe films, i.e. $\eta \approx 0.3 (0.17)$\cite{Nembach.2015}. However, we caution that such an approach is still limited by the $T^{3/2}$ dependence of $M_\mathrm{S}(T)$ derived for 3D systems. Overall, a generalization of the 3D Bloch law approach to ultrathin ferromagnets is inherently challenging. 

\paragraph{Bloch Law Revised for 2D}
A more intuitive approach has been proposed for ultrathin films, which explicitly incorporates their 2D character within the $M_\mathrm{S}(T)$ expression \cite{Yastremsky.2019}. As a first approximation, it appears sufficient to consider only the fundamental spin-wave mode, which has a uniform profile over the film thickness. The simplest dispersion for such a 2D model is given by $\omega(k) = \omega_0 + 2 \gamma A/M_\mathrm{S}(0) k^2$ \cite{Yastremsky.2019}. For this dispersion, $\rho(\omega)$ is a constant in 2D, proportional to $M_\mathrm{S}(0)/A$. Notably, the inclusion of a frequency gap $\omega_0$, due to dipolar interactions, anisotropies and external fields, is critical to model a finite $M_\mathrm{S}$ for $T>0$ \cite{Mermin.1966}. This results in the expression \cite{Yastremsky.2019}
\begin{equation}
	\label{eq:2D}
	M_\mathrm{S}(T)_{2D} = M_\mathrm{S}(0) \left[1- \frac{\upmu_\mathrm{B} k_\mathrm{B} T}{4 \pi \gamma A t \hbar} \ln \left( \frac{k_\mathrm{B} T}{\hbar \omega_0} \right) \right]\,.
\end{equation}
In this case (c.f. Bloch law), the determination of $A$ requires knowledge of the film thickness ($t$), but is independent of $M_\mathrm{S}(0)$.

\paragraph{2D Confinement Modes}
However, the 2D model needs considerable improvement if thickness modes are significantly populated \cite{Seavey.1958}. Even for nanometer thick films, the thermal population of perpendicular standing spin-wave (PSSW) modes can be considerable at temperatures typically usually used in magnetometry measurements. An estimate of their contribution can be obtained by considering the Bose-Einstein distribution $N(\omega)$ with
\begin{equation}
	\label{thermal_population}
	N(\omega_n) = \frac{1}{\exp\left(\frac{\hbar\omega_n}{k_\mathrm{B}T}\right)-1}\,  
\end{equation}
with the PSSW frequencies
\begin{equation}
	\label{PSSW_frequencies}
	\omega_\mathrm{n} =2 \gamma (A/M_\mathrm{S}) \cdot  \left(\frac{n\cdot \pi}{t} \right )^2\,
\end{equation}

\begin{figure*}
	\centering
	\includegraphics[width=0.95\textwidth]{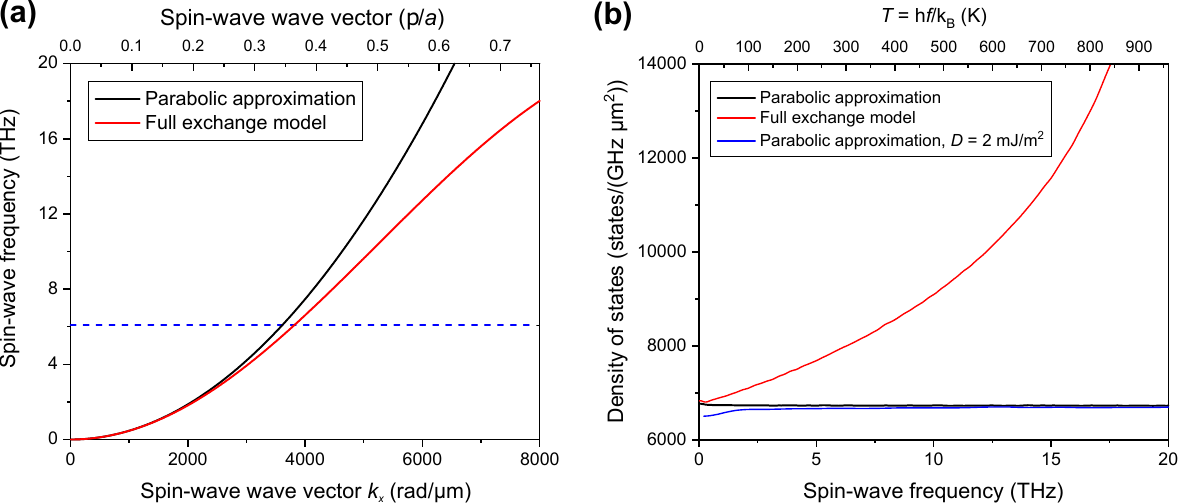}
	\caption{(a) Spin-wave dispersion for the full Heisenberg model (red, Eqn.~\ref{dispersion_full}) and its parabolic approximation (black,Eqn.\ref{dispersion_parabolic}) (model details in Appendix, parameters in main text). Dashed blue line denotes $f = k_\mathrm{B}T/h$ for $T = \SI{293}{K}$, corresponding to the maximum thermal population of spin wave states for this work. (b) Spin-wave DOS for different dispersion models in 2D: parabolic (black), full-exchange (red), and parabolic with DMI (blue, D  $= \SI{2}{mJ \per m^2}$) models.}
	\label{dispersion-deviation}
\end{figure*}

For $t = \SI{2}{nm}$, $M_\mathrm{S} = \SI{1200}{kA \per m}$, and $A = \SI{6}{pJ \per m}$, we estimate  the first PSSW mode with a wave vector component $k_\perp^{(1)} = \pi / t$ perpendicular to the film plane to be $\omega_1 \approx 2 \pi \cdot \SI{1.3}{\tera Hz}$. Likewise, the second PSSW mode ($k_\perp^{(2)} = 2 \pi / t$) $\omega_2 \approx 2 \pi \cdot \SI{5.5}{\tera Hz}$. Hence, assuming the fundamental mode, $\omega_0 \approx 2 \pi \cdot \SI{20}{\giga Hz}$, the relative occupation of the PSSW modes at RT is $N(\omega_1)/N(\omega_0)=\SI{1.3}{\%}$ and $N(\omega_2)/N(\omega_0)=\SI{0.2}{\%}$. Higher PSSW modes would have  considerably lower occupation. Thus, while it appears necessary to account for PSSW modes to adequately describe the $M_\mathrm{S}(T)$ reduction, the first two modes seem to be sufficient. Using instead a pure 2D model that neglects PSSW modes would underestimate the magnon density and, thereby, the $M_\mathrm{S}(T)$  reduction, ultimately resulting in an underestimate of $A$. 

\paragraph{2D Bloch Law with PSSW Modes}
The contribution of the thermal population of PSSW modes to the $M_\mathrm{S}(T)$ reduction can be described by models accounting individually for the influence of each mode \cite{Erickson.1991,Yastremsky.2019}. Considering the first three modes as per above, the $M_\mathrm{S}(T)$  reduction is given by  \cite{Erickson.1991,Yastremsky.2019}
\begin{equation}
	\label{m_reduction_pssw}
	\begin{split}
		M_\mathrm{S}(T)_\mathrm{PSSW}& \\
		= M_\mathrm{S}(0) &\left[ 1 - \sum_{n=0}^2 \frac{\upmu_\mathrm{B} k_\mathrm{B} T}{4 \pi \gamma A t \hbar} \ln{\left(\frac{\exp(\hbar \omega_n / k_\mathrm{B} T)}{\exp(\hbar \omega_n / k_\mathrm{B} T)-1}\right)} \right].
	\end{split}
\end{equation}
We note that the spin-wave dispersion deviates from the assumed parabolic shape towards the BZ boundary (compare Fig. \ref{dispersion-deviation}a)). Therefore, all approaches presented above are valid only at temperatures low enough that thermal excitations of high-$k$ magnon states can be neglected. This fact can be illustrated for the 2D case by examining the spin-wave DOS, which governs $M_\mathrm{S}(T)$ (see Fig. \ref{dispersion-deviation}(b)). In particular, Fig.~\ref{dispersion-deviation}(a) compares two spin-wave dispersion models for a 2D square lattice, calculated using $\omega_0 = 2 \pi \cdot \SI{10}{GHz}$, $M_\mathrm{S} = \SI{1200}{kA \per m}$, $A = \SI{10}{pJ \per m}$, and lattice constant $a = \SI{3}{\angstrom}$, consistent with typical ultrathin metallic ferromagnets \cite{Vaz.2008}. Fig.~\ref{dispersion-deviation}(b) evidences the implications of using the parabolic approximation to evaluate the spin-wave DOS for the Bloch law. While the DOS for the parabolic model is constant, that for the full dispersion model increases with frequency. Consequently, with increasing temperature, the full model populates more states of higher frequency, and due to its increasingly larger DOS, $M_\mathrm{S}(T)$ decreases faster than for the parabolic model. Therefore, any $M_\mathrm{S}(T)$ model assuming parabolic dispersion (e.g. Eqn. \ref{m_reduction_bloch}, \ref{eq:2D}, \ref{m_reduction_pssw}) underestimates the spin-wave DOS and, thereby the $M_\mathrm{S}(T)$  reduction, leading to a larger predicted $M_\mathrm{S}$ for a given $A$ and $T$. As a result, these models consistently underestimate $A$, which stabilizes ferromagnetic order against thermal fluctuations. 

\paragraph{Bloch Law: Notes}
Meanwhile, interfacial DMI has negligible influence on the DOS (Fig.~\ref{dispersion-deviation}(b), blue curve, $D = \SI{2}{mJ \per m^2}$). Hence, we expect that for $M_\mathrm{S}(T)$-based estimation of $A$ of ultrathin magnetic films from, both IP and OP measurements can be treated equally to good approximation. Finally, in the context of $M_\mathrm{S}(T)$  analysis, it is noteworthy that both the Bloch $T^{3/2}$ law, and the other models discussed here are formulated for $T$-independent exchange stiffness $A$.
\subsection*{Domain Periodicity}
%Extraction of $D/A$, extract $A$ using $D$ from BLS and DFT?
\paragraph{Domain Periodicity Principle}
In chiral multilayers, the microscopic domain characteristics are determined by the competition between $D$ and $A$. For multilayers exhibiting a labyrinthine domain state at remanence, it is established that the measured domain periodicity can be compared with micromagnetic simulations to determine the ratio of $A$ and $D$. An independent determination of $D$ can then be used to extract $A$ \cite{Soumyanarayanan.2017,MoreauLuchaire.2016, Kuepferling.2020}. A key limitation of this technique, especially for stacks lacking multiple repetitions, is that the required labyrinthine domain configuration is achievable only for a narrow range of magnetic layer thicknesses (e.g. 0.1--0.2~nm)  \cite{Boulle.2016}.
\subsection*{Density Functional Theory}
\paragraph{DFT: Principle & Method}
To complement the experimental results, density functional theory (DFT) calculations were performed to estimate $A$ \cite{Blaha.2020,Laskowski.2004} on appropriately constructed atomic multilayer slabs. The generalized Bloch’s theorem was used to generate spin spirals with wave vectors $k$ directed along the direction of the IP nearest neighbor atom, and the spiral axis is given by the cross product of the $k$-vector and slab normal with a spiral angle of \SI{90}{\degree} \cite{Laskowski.2004,Sandratskii.1998}. To extract the symmetric exchange $A$, the spiral energy density was calculated over a range of wave vectors $k$ and fit to a quadratic function. The accuracy of these methods have been assessed extensively in previous works, and shown to consistently predict trends similar to experiments \cite{ Soumyanarayanan.2017,Chen.2021}.

\section{Results}\label{Sec_Results}%%
\paragraph{Results Overview}
The values of $A$ extracted from the different techniques introduced in \S\ref{Sec_Methods} are plotted in Fig.~\ref{exchange_plot} and summarized in Tbl.~\ref{tab:summary}. Below, we introduce the samples studied in this work, and discuss the results obtained using the respective measurement techniques.

\subsection*{Samples}
%Summary, motivation for sample set, MS and anisotropy
\paragraph{Samples}
The multilayer thin film samples studied in this work were deposited on thermally oxidized Si wafers by DC magnetron sputtering at RT using a Chiron$^\text{TM}$ UHV system (base pressure $<$ \SI{5d-8}{torr}) from Bestec GmbH. 
Five sample compositions are examined, with ultrathin ($\leq 2$ nm) ferromagnetic (FM) layers: Pt/Co(1)/Pt, Ir/Co(1)/Pt, Ir/Fe(0.4)/Co(0.6)/Pt, Pt/Co(2)/Pt, Ir/Co(2)/Pt. The FM layer thicknesses, in nm, are indicated in parentheses, while all heavy metal layers are 1 nm thick. The stack additionally includes seed layers Ta(4)/Pt(5) for optimal texture, and a Pt(2) cap to protect against oxidation.
Of the five samples, only one (Pt/Co(2)/Pt) has a distinct IP easy axis. This sample set is designed to systematically compare the effects of inversion symmetry (symmetric Pt/Co/Pt c.f. the rest), FM bilayer (Ir/Fe(0.4)/Co(0.6)/Pt c.f. Ir/Co(1)/Pt), and varying FM thickness (Co(1) c.f. Co(2)) on $A$ and its experimental determination.

\subsection*{BLS Spectroscopy and Dispersion Analysis}

\begin{figure}
	\centering
	\includegraphics[width=0.42\textwidth]{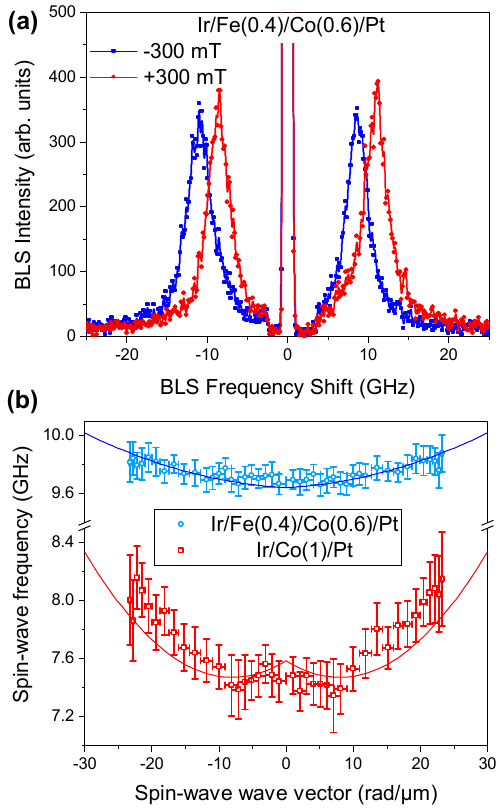}
	\caption{(a) Representative thermal BLS spectrum for Ir/Fe(0.4)/Fe(0.6)/Pt sample (incidence angle, $\varphi= \SI{45}{^\circ}$) at $\upmu_0 H_\mathrm{ext} = \pm$ \SI{300}{mT} (red, blue). (b) BLS spectroscopy measurement of the symmetric component of spin-wave dispersion, $f_\mathrm{sym}$ (from Eqn.~\eqref{Eq:Dispersion_sym}, data markers), and corresponding best fits (Eqn.\ref{Eq:Dispersion_sym}) used to estimate $A$ for two representative samples --  Ir/Fe(\SI{0.4}{nm})/Co(\SI{0.6}{nm})/Pt (blue, $\upmu_0 H_\mathrm{ext} = \SI{300}{mT}$) and Ir/Co(\SI{1}{nm})/Pt (red, $\upmu_0 H_\mathrm{ext} = \SI{830}{mT}$). Fit parameters are listed in Tbl.~\ref{tab:summary}.} 
	\label{BLS}
\end{figure}

\paragraph{BLS Data}
BLS measurements were performed using a wave-vector-resolved setup operated in the backscattering geometry \cite{Sebastian.2015}. Here, a laser with a wavelength of $\lambda_\mathrm{L} = \SI{532}{nm}$ was used and the spectral analysis of the inelastically scattered light was performed using a tandem-Fabry-Pérot interferometer of Sandercock-type \cite{Hillebrands.1999}.The measurements were performed at room temperature, with the external magnetic field applied IP to the magnetic film, with sufficiently large magnitude to saturate each sample in the in-plane direction, while oriented perpendicular
%The measurements were performed at room temperature, with the external magnetic field applied IP to the magnetic film -- to saturate the sample -- and perpendicular 
to the incident plane of the probing laser light. Wave vector resolution can be achieved by varying the angle of incidence, $\varphi$, such that the probed spin-wave wave vector equals $k = 4 \pi \sin(\varphi) / \lambda_\mathrm{L}$ \cite{Sebastian.2015}. For the IP configuration, BLS probes thermally populated magnetostatic surface spin waves described by Eqn.~\eqref{Eq:Dispersion_sym}, which propagate perpendicular to the applied (IP) field. Fig.~\ref{BLS}(a) shows representative BLS spectra for a chiral multilayer sample. The characteristic asymmetry in the (anti) Stokes spin wave peaks for opposite field polarities arises from the chiral DMI interaction. For this work, however, we are mainly interested in the symmetric component of the dispersion. Accordingly, Fig.~\ref{BLS}(b) shows representative symmetrized BLS dispersion data obtained for two samples, with the fitted result from the dispersion model (Eqn.~\ref{Eq:Dispersion_sym}) overlaid. This data is obtained by comparing the measurements at opposite directions of the magnetic bias field.

\paragraph{BLS Fitting}
We note from Eqn.~\ref{Eq:Dispersion_sym} that the dispersion curvature is also influenced  by the anisotropy field, $H_\mathrm{U}$ and the saturation magnetisation $M_\mathrm{S}$. Hence, the fitted value for $A$ can be influenced by $\upmu_0 H_\mathrm{U}$, which in turn influences the optimal $M_\mathrm{S}$ since these two parameters essentially define the ferromagnetic resonance frequency $f(k\rightarrow 0)$ (FMR). As a basis for the evaluation of the BLS data, we use the  $H_\mathrm{U}$ and $M_\mathrm{S}$ values obtained from SQUID measurements. To estimate the error bar for $A$ caused by the uncertainty of $M_\mathrm{S}$ and $H_\mathrm{U}$  , we allow $A$ to vary in either direction until $M_\mathrm{S}$ needs to altered by $\sim$\SI{15}{\%} -- corresponding to the uncertainty of $M_\mathrm{S}$ within VSM measurements. To get a self-consistent modelling, for every $M_\mathrm{S}$ value, we adjust $H_\mathrm{U}$ to match the calculated FMR frequency to the frequency measured by BLS for $k\rightarrow 0$.

\paragraph{BLS Trends}
For the Ir/Fe(\SI{0.4}{})/Co(\SI{0.6}{})/Pt sample, the fit thus obtained agrees well with the data for $A = \SI{7}{pJ \per m}$, and $\upmu_0 H_\mathrm{U} = \SI{1433}{mT}$ ($\upmu_0 H_\mathrm{k} \approx \SI{120}{mT}$). Meanwhile, for the Ir/Co(\SI{1}{nm})/Pt sample, where the dispersion data shows a much larger slope, the best fit model parameters are $A = \SI{22}{pJ \per m}$ and $\upmu_0 H_\mathrm{U} = \SI{2210}{mT}$ ($\upmu_0 H_\mathrm{k} \approx \SI{730}{mT}$). In general, we find a good agreement between the $H_\mathrm{k}$ values obtained from BLS with those from VSM as well as with previous studies on similar material systems \cite{Soumyanarayanan.2017}. 
The values of $A$ obtained using Eqn.~\ref{Eq:Dispersion_sym} on the BLS data for all other samples are summarized in Fig.~\ref{exchange_plot}. We verified that more sophisticated modelling of the dispersion relation (see Appendix) using a layer resolved numerical approach which allows, e.g. for a distribution of material parameter across the film thickness, gives values for $A$ which are in agreement with those obtained from Eq.~\ref{Eq:Dispersion_sym} within the error bars. Note that the error bars for $A$ using BLS are comparatively large, as it requires the experimental determination of several parameters, each using different techniques, and with different dependencies. A detailed list of values for the relevant material parameters for all samples is provided in Tbl.~\ref{tab:summary}.

%The error bars given for $A$ from the dispersion analysis have been estimated as follows: First, it needs to be pointed out that both the symmetric exchange and the anisotropy field, $\upmu_0 H_\mathrm{U}$ affect the curvature of the dispersion relation. Hence, we vary the symmetric exchange $A$ and balance the resulting modification of the curvature of the dispersion relation by a modification of the anisotropy field. Since the FMR frequency ($\omega_0$) sensitively depends on the anisotropy field, this requires an adaption of the saturation magnetization, $M_\mathrm{S}$. We allow for a variation of the symmetric exchange until a modification of $M_\mathrm{S}$ of about \SI{15}{\%} is reached corresponding to the uncertainty of the VSM measurements used to determine $M_\mathrm{S}$ which then gives the limits for the value of the symmetric exchange. 
% the measurement has been performed at an applied field of $\upmu_0 H_\mathrm{ext} = \SI{830}{mT}$. 

\subsection*{\texorpdfstring{\textit{M}\textsubscript{S}(\textit{T}) Results and Analysis}{M(T) Results and Analysis}}
\begin{figure}
	\centering
	\includegraphics[width=0.48\textwidth]{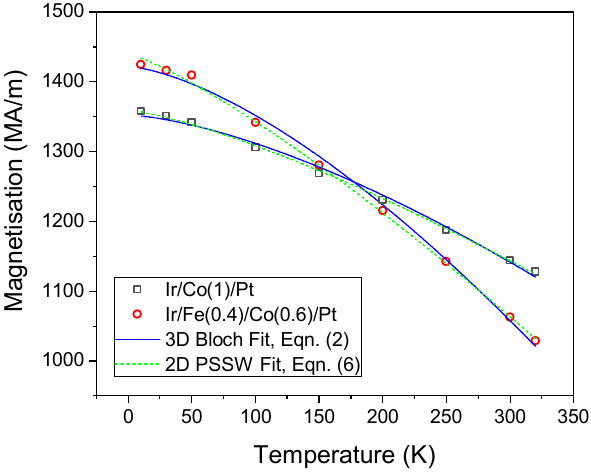}
	\caption{$T$-dependence of saturation magnetization, $M_{S}$, for samples Ir/Co(\SI{1})/Pt (black squares) and Ir/Fe(\SI{0.4})/Co(\SI{0.6})/Pt (red dots) obtained from the zero field extrapolation of $M(H)$ hysteresis loops measured by SQUID magnetometry. Lines denote fits to Bloch's $T^{3/2}$ law used to obtain $A$ via the 3D model (solid blue, Eqn.~\ref{m_reduction_bloch}) and the 2D PSSW model (dashed green, Eqn.~\eqref{m_reduction_pssw}, third iteration).}
	\label{M_T}
\end{figure}

\paragraph{MT Data}
Fig.~\ref{M_T} shows representative $M_\mathrm{S}(T)$ data obtained from the same two samples -- Ir/Fe(\SI{0.4}{})/Co(\SI{0.6}{})/Pt and Ir/Co(1)/Pt -- from $M(H)$ hysteresis loops obtained by SQUID magnetometry measurements at varying temperatures in the IP configuration. The measured $M_\mathrm{S}(T)$ data were fit with Bloch's $T^{3/2}$ law using: (a) the 3D value for $\eta$, (b) the thickness corrected $\eta$ values \cite{Nembach.2015}, and (c) the model including PSSW modes (Eqn.~\eqref{m_reduction_pssw}), as shown in Fig.~\ref{M_T}. To fit Eqn.~\eqref{m_reduction_pssw}, $\omega_0$ was first estimated using the midpoint of the used $M(H)$ field range, i.e., $\omega_0 = \gamma \upmu_0 (H_\mathrm{ext} + H_\mathrm{U} - M_\mathrm{S})$ \cite{Kittel.1948}, and was then used in Eqn.~(\ref{m_reduction_pssw}) to estimate $A$ using an iterative fitting procedure (see Appendix for details). Here, $M_\mathrm{S}$ was estimated by extrapolating  $M_\mathrm{S}(T)$ data to $T=0$, while $H_\mathrm{U}$ was extracted from BLS measurements modelled by Eqn.~\ref{Eq:Dispersion_sym}). 

%In particular, we also find good agreement between the two SQUID measurement methods with the resulting values for the exchange stiffness deviating by less than \SI{6}{\%}, such that we claim that both these methods can be employed to obtain reliable results. A further elaboration of the results from specific measurement configurations can be found in the Appendix.

%With the field oriented along the film normal, any DMI-induced non-reciprocity in the spin-wave dispersion relation can be excluded, which is the basis of the $M_\mathrm{S}(T)$ models in Eqs.~\eqref{m_reduction_bloch} and \eqref{m_reduction_pssw}. Nevertheless, as described above and in the Appendix, the influence of the DMI on the spin-wave dispersion relation for the in-plane field configuration does not significantly alter the results, as we have experimentally verified.

\paragraph{MT Trends}
The values of $A$ obtained from the various $M_\mathrm{S}(T)$ models are plotted in Fig.~\ref{exchange_plot}, and detailed in Tbl.~\ref{tab:summary}. We find that the modified $\eta$ version of Bloch's law ($\eta \approx 0.3 (0.17)$ for $t=\SI{1 (2)}{nm}$ \cite{Nembach.2015}) returns the highest values for $A$, providing the best agreement with those obtained from BLS and DFT. Incidentally, the $M_\mathrm{S}(T)$ measurements were repeated in the OP configuration, with no observable differences in the resulting $A$ values c.f. IP results, in line with the expected negligible influence of DMI on the spin-wave DOS (c.f. Fig.~\ref{dispersion-deviation}(b)). Separately, we also performed direct measurements of the $T$-dependence of magnetization, i.e. $M(T)$ at fixed applied fields above saturation ($H > H_\mathrm{S}$). While this latter approach allows for a more precise estimation of $\omega_0$ in Eqn.~\eqref{m_reduction_pssw}, we find excellent agreement of the obtained $A$ values with the hysteresis loop method. A detailed description of these quantitative comparisons is provided in the Appendix.
%Clearly, Bloch's $T^{3/2}$ law leads to the lowest exchange values while the discrepancy between results from the model including PSSW modes and other experimental techniques is smaller.
%Calculation of PSSW frequencies: Use M_S value from extrapolation towards zero T (since M(T) analysis gives A at zero temperature)

\subsection*{Domain Periodicity}
\begin{figure}
	\centering
	\includegraphics[width=0.48\textwidth]{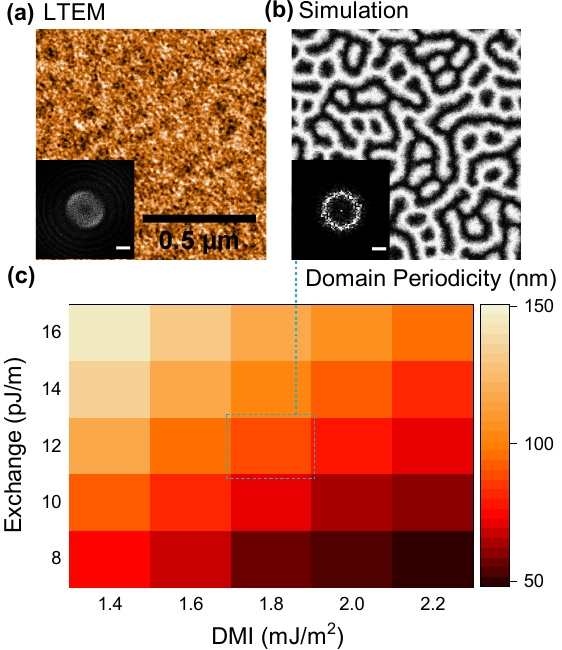}
	\caption{Zero field domain periodicity analysis for sample [Ir/Fe(0.4)/Co(0.6)/Pt]$_2$. (a) Lorentz TEM image and (b) micromagnetically simulated magnetization ($M_z$) with $D$, $A$ parameters closest to the best fit (scalebar: 0.5~$\upmu$m). Insets show Fourier transforms (scalebar: \SI{10}{\upmu m^{-1}}). (c) Simulated domain periodicity for an array of $D$, $A$ values. Dashed box shows the closest match to experiment (i.e., b).}
	\label{DA_matrix}
\end{figure}%

\paragraph{Periodicity Constraints & Results}
Domain periodicity determination of $A$ requires imaging of magnetic texture configuration at remanence. Of the five samples studied, only  Ir/Fe(0.4)/Co(0.6)/Pt can stabilize a ZF domain configuration  (Fig.~\ref{DA_matrix}(a)). As a result domain periodicity analysis could not be performed on the other four samples -- for  Pt/Co(\SI{1}{})/Pt and Ir/Co(\SI{1}{})/Pt, because of their high remnant magnetization, and for Co(2) samples, due to their IP easy axis -- both of which result in the lack of domain nucleation at remanence. 

\paragraph{Domain Periodicity Method}
The domain imaging was performed using Lorentz transmission electron microscopy (LTEM) with an FEI Titan S/TEM operated in Fresnel mode at 300~kV. A dedicated Lorentz lens, used to focus the electron beam, was used at a defocus of $-1.8$~mm, while the objective lens located at the sample position was switched off for field-free image acquisition. In order to obtain sufficient magnetic contrast for LTEM, a two-repeat version of the stack, i.e. [Ir/Fe(0.4)/Co(0.6)/Pt]$_2$, was used. Despite the weak signal in real space, the signature `split-ring' structure of the labyrinthine domain configuration is clearly visible in frequency space, which allows us to obtain a domain period of \SI{87}{nm}. 
Micromagnetic simulations were performed using MuMax$^3$ \cite{Vansteenkiste.2014} for an array of $D$ and $A$ values (\ref{DA_matrix}(b,c)), with $M_\mathrm{S}$ and uniaxial anisotropy ($K_\mathrm{u}$) parameters obtained from magnetometry. Subsequently, the real space magnetization of each simulation was Fourier-transformed to extract the period ($P$), which was fit to a low-order 2D polynomial: $P = a_0 + a_1 D + a_2 A + a_3 D \cdot A$ ($R^2 = 0.99$). 
By constraining the fit to the measured period (87~nm) and $D = \SI{1.72}{mJ \per m^2}$ (measured by BLS), $A$ was determined to be $\SI{11.1}{pJ \per m}$. 

\subsection*{DFT Calculations}
\begin{figure}
	\centering
	\includegraphics[width=0.48\textwidth]{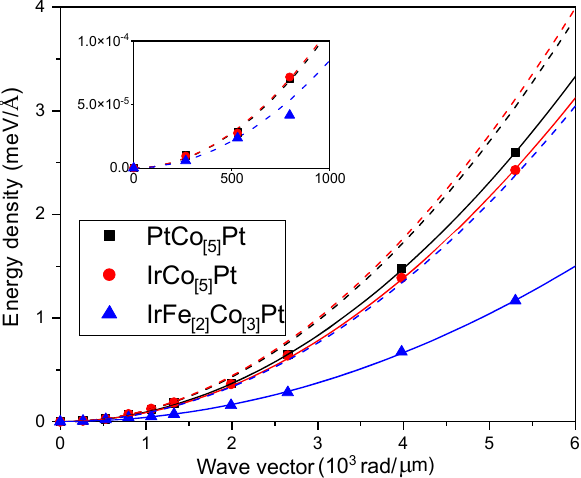}
	\caption{DFT-computed dispersion of spin spiral energy densities for the three 1~nm samples studied in this work. Inset shows a zoom-in view of the small-$k$ region ($k$ axis in {rad}{/\textmu m}). Dashed (solid) lines are quadratic fits for small (large) $k$ values, respectively.}
	\label{dft}
\end{figure}%

\paragraph{DFT Dispersion}
DFT calculations were implemented on atomic multilayer slabs with compositions PtCo$_{[5]}$Pt, IrCo$_{[5]}$Pt, IrFe$_{[2]}$Co$_{[3]}$Pt, PtCo$_{[9]}$Pt, and IrCo$_{[9]}$Pt, where subscripts (in brackets) for Fe and Co represent the number of atomic layers. As each Fe (Co) atomic layer is about \SI{0.2}{nm} thick, these slabs approximately correspond to the five experimentally studied samples.
%DFT calculations were implemented on atomic multilayer slabs with compositions PtCo$_{[5]}$Pt, IrCo$_{[5]}$Pt, and IrFe$_{[2]}$Co$_{[3]}$Pt, where subscripts (in brackets) for Fe and Co represent the number of atomic layers. As each Fe (Co) atomic layer is about \SI{0.2}{nm} thick, these slabs correspond to experimental thicknesses of Pt/Co(1)/Pt, Ir/Co(1)/Pt, and Ir/Fe(0.4)/Co(0.6)/Pt. 
Each slab is separated by a vacuum of \SI{10}{\angstrom} in the normal direction to prevent spurious inter-cell interactions, the IP lattice constant is set to the bulk Ir value, and exchange-correlation approximated by the Perdew-Burke-Ernzerhof formulation of the generalized gradient approximation \cite{Perdew.1996}.
The energy densities for spin spiral configurations were calculated for wave vectors $k$~up~to~$\sim$ 6000 {rad}{/\textmu m} (0.6 {rad/\AA}), as shown in Fig.~\ref{dft} (see Methods for details). The spiral energy shows approximately quadratic dependence on $k$, and low wave vectors, and progressively deviates from parabolic behavior at large $k$. For the IrFe$_{[2]}$Co$_{[3]}$Pt case, we additionally note appreciable softening of the spin spiral at $k \sim$ 800 {rad}{/\textmu m} (0.08 {rad/\AA}). This phenomenon has been observed in IrFe slabs, and attributed to Ir-Fe hybridization \cite{Bergmann.2006}. 

\begin{figure}
	\centering
	\includegraphics[width=0.48\textwidth]{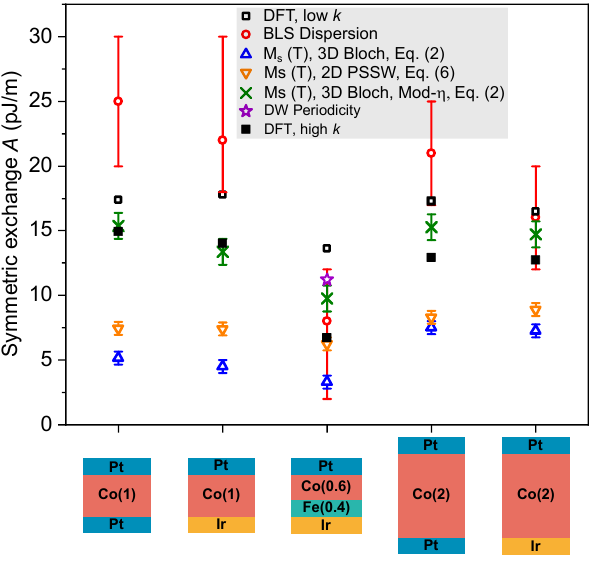}
	\caption{ Exchange stiffness, $A$ of the five studied samples, obtained by different methods detailed -- BLS dispersion, $M_\mathrm{S}(T)$ modeling (3D Bloch law, modified $\eta$, PSSW model), domain periodicity, and DFT calculations (low $k$, high $k$).
		Bottom inset shows stack schematics for the respective samples, dashed black line separates the 1~nm FM and 2~nm FM samples for clarity.
	}\label{exchange_plot}
\end{figure}

\paragraph{DFT Trends}
To determine $A$, we fit the spiral energy densities to a quadratic function in $k$ over varying ranges of \textit{k}, which yield different results due to the deviation from parabolicity. For the small $k$ limit, we fit to energies with $k <$1200 {rad}{/\textmu m} (0.12 {rad/\AA}), while for large $k$, we fit over all the data shown in Fig.~\ref{dft}.
%For the small $k$ limit, we fit to energies with $k <$1200 {rad}{/\textmu m} (0.12 {rad/\AA}), yielding $A$ values of 17.4, 17.8 and 13.6 pJ/m respectively. Meanwhile for large $k$, we fit over all the data shown in Fig.~\ref{dft}, which gives $A$ values of 14.9, 14.0 and 6.7~pJ/m respectively.
Overall, the DFT-computed results, shown in Fig.~\ref{exchange_plot} and Tbl.~\ref{tab:summary}, give the largest exchange for slabs with pure Co layers, and lower with the introduction of Fe. Small variations are observed with varying heavy metal layers (Pt/Co/Pt c.f. Ir/Co/Pt) and Co thickness (1 nm Co c.f. 2 nm Co).
%Overall, the DFT-computed results, shown in Fig.~\ref{exchange_plot} and Tbl.~\ref{tab:summary}, give the largest exchange for slabs with pure Co layers, and lower with the introduction of Fe. Meanwhile, little variation is observed with varying heavy metal layers (PtCo$_{[5]}$Pt c.f. IrCo$_{[5]}$Pt).
Finally, the $A$ obtained for small $k$ is consistently larger than that obtained over a larger $k$ range -- most notably by $\sim2\times$ for IrFe$_{[2]}$Co$_{[3]}$Pt. This difference is reasonable considering that the spin spiral dispersion flattens near the BZ boundaries, and is relevant to the observed discrepancies in $A$ between BLS and $M_\mathrm{S}(T)$ measurements (see \S\ref{Sec_Discussion}). 

\section{Discussion}\label{Sec_Discussion}%%
%%%%% Section D: Discussions

%% D1: Quantitative Comparison of Methods
\paragraph{Trends Summary}
The key result of this work, presented in Fig.~\ref{exchange_plot} and Tbl.~\ref{tab:summary}, compares the $A$ values of five multilayer samples obtained using the different methods detailed above. Overall, we find that the $A$ values from different methods do not coincide within errors bars for any sample, with up to $2-5\times$ discrepancies within each sample across techniques. We conclude that these deviations between techniques are likely of systematic nature, and may arise from the specific assumptions and limitations of the methods for ultrathin ferromagnets, discussed further below. As the first cross-technique comparison of exchange stiffness in ultrathin FMs to our knowledge, our work may offer a viable explanation of the large spread of $A$ values reported on similar ultrathin films \cite{Metaxas.2007,Vaz.2008,Shepley.2018,Yastremsky.2019,Shahbazi.2019}.

%% D2: Sample Trend Across Methods
\paragraph{Sample Trend Across Methods}
Reassuringly however, a qualitative comparison of techniques reveals that the evolution of $A$ across samples with the same FM thickness follows a similar trend.
For example, for the 1~nm thick FM samples, all methods report $A$ values for Pt/Co(1)/Pt that are comparable or larger than that for Ir/Co(1)/Pt, while that for Ir/Fe(0.4)/Co(0.6)/Pt is considerably lower. The decrease in $A$ upon introducing Fe, also noted by DFT, is consistent with the lower exchange for bulk Fe compared to bulk Co \cite{Pajda.2001}. Additionally, interfacial hybridization effects, which assume increased significance for ultrathin films, may also play a role in the trend observed across the 1 nm and 2nm FM samples \cite{Kim.2018, Perini.2018, Zakeri.2013}.

%For example, for the 1~nm thick FM samples, all methods report the highest $A$ for Pt/Co(1)/Pt, followed by Ir/Co(1)/Pt, while that for Ir/Fe(0.4)/Co(0.6)/Pt is considerably lower. The decrease in $A$ upon introducing Fe, also noted by DFT, is consistent with the lower exchange coupling for bulk Fe compared to bulk Co \cite{Pajda.2001}.%

%% D3: Method Trend for Samples P1
\paragraph{Method Trend for Samples: BLS}
Meanwhile, comparing across techniques for any given sample, we find that BLS analysis consistently reports the largest magnitude for $A$. Note that the the large error bars for BLS in Fig.~\ref{exchange_plot} reflect the inherent challenges in the determination of $A$ from BLS spin-wave spectra for ultrathin films with strong OP anisotropy, due to the intermixing of the different contributions in Eqn.~\ref{Eq:Dispersion_sym}. However, even within their error bars, most  $A$ values from BLS are considerably larger than the ones obtained from other methods. As an interesting aside, we note that recent comparative measurements of the DMI, $D$, also reported higher values from BLS measurements than other methods like domain wall expansion \cite{Magni.2022}.
\begin{table*}[htb]
	\renewcommand{\arraystretch}{2}
	\begin{ruledtabular}
		\begin{tabular}{c|c|ccccc}
			\bf{Parameter} & \,\bf{Technique}\, &
			\parbox[c]{2cm}{Pt/Co(1)/Pt}& 
			\parbox[c]{2cm}{Ir/Co(1)/Pt} & 
			\parbox[c]{3.15cm}{Ir/Fe(0.4)/Co(0.6)/Pt} & 
			\parbox[c]{2cm}{Pt/Co(2)/Pt} & 
			\parbox[c]{2cm}{Ir/Co(2)/Pt}\\  
			\hline\hline
			$M_{\mathrm{S}}\,(\SI{}{kA\per m})$ & \multirow{3}{*}{VSM } & $\SI{1430\pm220}{}$  & $\SI{1280\pm190}{}$  & $\SI{1200\pm180}{}$  & $\SI{1330\pm200}{}$  & $\SI{1200\pm190}{}$ 
			\tabularnewline
			\cline{1-1} \cline{3-7} \cline{4-7} \cline{5-7} \cline{6-7} \cline{7-7} 
			$K_{\mathrm{U}}\,(\SI{}{MJ\per m^{3}})$  &  & $\SI{1.93\pm0.29}{}$  & $\SI{1.54\pm0.23}{}$  & $\SI{0.86\pm0.13}{}$  & $\SI{0.81\pm0.12}{}$  & $\SI{0.89\pm0.13}{}$ 
			\tabularnewline
			\cline{1-1} \cline{3-7} \cline{4-7} \cline{5-7} \cline{6-7} \cline{7-7} 
			$K_{\mathrm{eff}}\,(\SI{}{MJ\per m^{3}})$  &  & $\SI{0.63\pm0.09}{}$  & $\SI{0.50\pm0.08}{}$  & $\SI{-0.05\pm0.05}{}$  & $\SI{-0.12\pm0.05}{}$  & $\SI{0.24\pm0.05}{}$ 
			\tabularnewline
			\hline 
			$\gamma\,(\SI{}{rad\per(T\cdot ns)})$  & FMR  & $\SI{184.96\pm0.87}{}^{\ast}$  & $\SI{167.50\pm0.13}{}^{\ast}$  & $\SI{167.88\pm0.19}{}$  & $\SI{184.96\pm0.87}{}$  & $\SI{167.50\pm0.13}{}$ \tabularnewline
			\hline 
			\multirow{4}{*}{$A\,(\SI{}{pJ\per m})$} & BLS  & $\SI{25\pm5}{}$  & $\SI{22}{}+8/-4$  & $\SI{7}{}+8/-4$  & $\SI{21\pm4}{}$  & $\SI{16\pm4}{}$ 
			\tabularnewline
			\cline{2-7} \cline{3-7} \cline{4-7} \cline{5-7} \cline{6-7} \cline{7-7} 
			& $M_{{\rm {S}}}(T)$: $T^{3/2}$, Eqn.\,\ref{m_reduction_bloch}  & $\SI{5.18\pm0.50}{}$  & $\SI{4.51\pm0.50}{}$  & $\SI{3.3\pm0.50}{}$  & $\SI{7.52\pm0.50}{}$  & $\SI{7.25\pm0.50}{}$ 
			\tabularnewline
			\cline{2-7} \cline{3-7} \cline{4-7} \cline{5-7} \cline{6-7} \cline{7-7} 
			& $M_{{\rm {S}}}(T)$: PSSW, Eqn.\,\ref{m_reduction_pssw}  & $\SI{7.45\pm0.51}{}$  & $\SI{7.41\pm0.23}{}$  & $\SI{6.26\pm0.24}{}$  & $\SI{8.33\pm0.32}{}$  & $\SI{8.91\pm0.35}{}$ 
			\tabularnewline
			\cline{2-7} \cline{3-7} \cline{4-7} \cline{5-7} \cline{6-7} \cline{7-7} 
			& $M_{{\rm {S}}}(T):T_{{\rm\bf{mod\,\eta}}}^{3/2}$, Eqn.\,\ref{m_reduction_bloch}  & $\SI{15.38\pm1.00}{}$  & $\SI{13.39\pm1.00}{}$  & $\SI{9.78\pm1.00}{}$  & $\SI{15.27\pm1.00}{}$  & $\SI{14.73\pm1.00}{}$ 
			\tabularnewline
			\cline{2-7} \cline{3-7} \cline{4-7} \cline{5-7} \cline{6-7} \cline{7-7} 
			& LTEM, Periodicity & - & - & $\SI{11.1}{}$  & - & -\tabularnewline
			\cline{2-7} \cline{3-7} \cline{4-7} \cline{5-7} \cline{6-7} \cline{7-7} 
			& DFT, low $k$  & $\SI{17.4}{}$  & $\SI{17.8}{}$  & $\SI{13.6}{}$  & $\SI{17.3}{}$  &  $\SI{16.5}{}$
			\tabularnewline
			\cline{2-7} \cline{3-7} \cline{4-7} \cline{5-7} \cline{6-7} \cline{7-7} 
			& DFT, high $k$  & $\SI{14.9}{}$  & $\SI{14.0}{}$  & $\SI{6.7}{}$  & $\SI{12.9}{}$  &  $\SI{12.7}{}$
			\tabularnewline
	    \end{tabular}
	\end{ruledtabular}
	\caption{\label{tab:summary}
		Measured values of the key magnetic parameters for the five studied samples, and the measurement techniques used. VSM was used to determine $M_\mathrm{S}$ (OP), and anisotropy parameters $K_\mathrm{U}$ and $K_\mathrm{eff}$. The $K_\mathrm{U}$ and $K_\mathrm{eff}$ obtained from BLS deviate from these values by less than 10\%. FMR is used to obtain $\gamma$ (Fig.~\ref{FMR}, $\ast$ Pt/Co(1)/Pt, Ir/Co(1)/Pt assumed to have the same $\gamma$ as measured for their 2~nm thick counterparts.). 
		The values of exchange stiffness, $A$, obtained by BLS dispersion, $M_\mathrm{S}(T)$ modeling (3D Bloch law, modified $\eta$, PSSW model), LTEM domain periodicity, and DFT calculations (low $k$, high $k$) are listed for comparison (methods and parameters detailed in \S\ref{Sec_Results}).}
\renewcommand{\arraystretch}{1}%
\end{table*}
%
%% D4: Method Trend for Samples P2
\paragraph{Method Trend for Samples: Others}
The values obtained from BLS for $A$ are followed by Bloch law (modified $\eta$), and the 2D PSSW Bloch law (Eqn.~\eqref{m_reduction_pssw}). Finally, the original 3D Bloch $T^{3/2}$ law (Eqn.~\eqref{m_reduction_bloch}) reports the smallest value of $A$ for the studied ultrathin films. 
In particular, the $A$ values from the 3D Bloch law are unrealistically low as it does not account for the increased magnon density in ultrathin films \cite{Nembach.2015}. Meanwhile, we expect that the 2D Bloch law, with the iterative use of Eqn.~\eqref{m_reduction_pssw}, should produce closer to those obtained from DFT and other measurement techniques. 
While this is indeed qualitatively the case, the $A$ values from the 2D Bloch law are still considerably smaller than those from the BLS dispersion. Finally, a similar trend is observed for the Bloch $T^{3/2}$ law with thickness-corrected $\eta$, albeit the $A$ values for this case are closer to the BLS values.

%% D5: Ms(T) Limitations: Non-Parabolic Dispersion
\paragraph{Ms(T) Limitation: Non-Parabolic Dispersion}
The fact that all $M_\mathrm{S}(T)$-based methods report considerably lower $A$ values compared to BLS could be attributed to several reasons. One likely source is the assumed parabolic dispersion of the spin wave spectrum which forms the basis for all techniques used to estimate $A$ -- albeit over different ranges of wave vectors. As shown by DFT calculations (Fig. \ref{dft}), the dispersion softening at higher $k$ results in discrepancies between parabolic fits over different ranges of wave vectors, leading to over $2\times$ variation in the obtained magnitude of $A$. 
On one hand, BLS probes the low $k$ limit ($k \lesssim$ 25 {rad}{/\textmu m}), where the parabolic approximation is expected to hold (Fig. \ref{dft}: inset). On the other hand, $M_\mathrm{S}(T)$-based methods probe a much larger $k$ range of thermal spin wave distribution, which may extend beyond the parabolic dispersion region. Therefore, as seen in Fig. \ref{dft}, a parabolic fit to the wave vector range probed by $M_\mathrm{S}(T)$-based methods would generally lead to an underestimation of $A$ in comparison to, e.g., the full Heisenberg model. 
Importantly, we caution that for most cases, a simple parabolic approximation may not be justified over the wave vector range probed by the $M_\mathrm{S}(T)$-based methods, and the values for $A$ thus obtained should be interpreted with care.
% In our particular case, also the DFT calculations clearly show that the $A$ values obtained from fits to the low $q$ part lead to significantly higher values for $A$ in comparison to an analysis including the high $q$ part.

%% D6: Ms(T) Limitations: PSSW Modes
\paragraph{Ms(T) Limitation: PSSW Modes}
Another aspect to note is that only within the $M_\mathrm{S}(T)$ measurements, the higher order standing spin-wave (PSSW) modes are also indirectly probed.
Here, we have modeled the PSSW modes using micromagnetic continuum theory (Eqn.~\eqref{m_reduction_pssw}), which approaches its limits for the nanometer thick FMs studied here. A recent work suggests that PSSW modes in ultrathin films may be renormalized to lower energies along the OP direction, while retaining their dispersion for the IP direction \cite{Pelliciari.2021}. To test this effect, we halve the frequencies of the PSSW modes compared to Eqn. \ref{PSSW_frequencies} for Ir/Co(1)/Pt within the 2D PSSW model (Eqn.~\ref{m_reduction_pssw}), in line with the renormalization factor found in \cite{Pelliciari.2021}. We find that this results in an increase of the modelled exchange stiffness by about 20 \%  compared to Eqn.~\ref{PSSW_frequencies}, which uses the micromagnetic continuum approach. This suggests that detailed modeling of the PSSW mode characteristics can at least partially account for the observed reduction of $A$ from $M_\mathrm{S}(T)$ methods compared to BLS measurements.

%% D7: Ms(T) Limitations: Single Particle Model & 
\paragraph{Ms(T) Limitation: Other Factors}
The accuracy of $M_\mathrm{S}(T)$ modeling may be further improved by also incorporating single particle excitations, as well as higher-order exchange interactions. On one hand, single particle excitations would increase the $M_\mathrm{S}(T)$ reduction, leading to an underestimate of $A$ \cite{Maeda.1973}. Therefore, incorporating these would reduce the discrepancy of $A$ c.f. BLS results. On the other hand, higher-order exchange interactions may be able to capture the more complex wave vector dependence of the exchange energy density \cite{Banerjee.2014b,Gutzeit.2021}. However, incorporating these would drastically increase the resources needed to extend the respective models, and may complicate the extraction of the Heisenberg exchange contribution.

%% D8: Discounting Dead Layer Effects
\paragraph{Discounting Dead Layer Effects}
%Another source of discrepancy to be considered is the potential presence of a dead layer within the magnetic film. This would affect
Other sources of discrepancy to be considered are the potential presence of a dead layer within the magnetic film, and proximity-induced magnetization effects within the neighbouring heavy metals. Both of these would the effective FM thickness, and therefore directly influence the results from all three $M_\mathrm{S}(T)$ models. The FM thickness also governs the dipolar interaction in the spin-wave dispersion relation, leading to an additional uncertainty. 
However, seeing as the measured $M_\mathrm{S}$ for these ultrathin samples is typically around \SI{1200}{kA \per m} at RT, largely in line with reported values for similar stacks \cite{Tan.2021}, we expect the net contribution of dead layer and magnetic proximity effects \cite{Masgrau.2015} to be negligible in this case.

%% D9: Temperature Dependence
\paragraph{$T$-Dependent Effects}
In this work, we have refrained from any $T$-dependent renormalization of the $A$-values obtained from $M(T)$ measurements. In principle, the need for such renormalization arises as the $M_{\rm s}(T)$ models putatively estimate $A$ for $T=0$, while the BLS-measured $A$ value is for room temperature. Some previous works have implemented such renormalization using \emph{ab initio} electronic structure calculations of bulk Co to relate the decrease of $A(T)$ with temperature to that of $M(T)$ , and to thereby obtain a scaling law for $A(M(T))$ \cite{Moreno.2016}. However, our work has not applied any such $T$-dependent renormalization to our $M(T)$ analysis for several reasons. First, an accurate renormalization treatment would, in principle, require similar \emph{ab initio} calculations for each of our thin film samples, as their electronic structure may differ considerably from the previously considered bulk Co case \cite{Moreno.2016}, which does not account for finite thickness and interface effects. Second, and more importantly, the derivation of Bloch's law and all its 2D variants explicitly assume the constancy of $A$ over the measured temperature range. Therefore, the inclusion of $T$-dependence of $A$ within the measured temperature range is inconsistent with the use of Bloch's law in its current form. Moreover, we further note that the approximation of a $T$-independent $A$ fits the measured $M(T)$ data  up to room temperature with very high accuracy. Thus, even if the present form of Bloch’s law is revised to incorporate $T$-dependence of $A$, such a fit may be expected to instead over-parametrize the problem, resulting in potentially spurious estimates. Finally, we emphasize that any $T$-dependent renormalization \cite{Moreno.2016} would lower the $A$-values resulting from $M(T)$ measurements. Thus, the lack of such renormalization cannot account for our findings of consistently lower $A$-values from $M(T)$ measurements as compared to other methods.

\section{Conclusion}\label{Sec_Conclusion}

%% Summary of Results
\paragraph{Results Summary}
In summary, this work presents a quantitative comparison of the determination of the Heisenberg exchange stiffness, $A$, across five multilayer films comprising ultrathin ferromagnets using different methods - viz. BLS dispersion, $M_\mathrm{S}(T)$ reduction via 3 models (conventional 3D Bloch law, 3D Bloch law with modified $\eta$, 2D Bloch law with PSSW modes), domain periodicity, and DFT calculations. Despite the exchange interaction being arguably the most fundamental property of magnets, and its crucial role in determining material viability for device applications, we find that a thorough understanding of its magnitude in the ultrathin film limit and its dependence on interfacial and stack properties is lacking. 

\paragraph{Explanation of Results}
While qualitative trends are apparent across samples, we note $2-5\times$ discrepancies in the magnitude of $A$ obtained by the various methods, underscoring the complexity of its determination in the ultrathin limit. In general, methods using $M_\mathrm{S}(T)$ models report lower $A$ values compared to BLS spin-wave dispersion analysis. Qualitatively, these discrepancies may arise, e.g., from the different wave vector ranges probed by these methods, and the deviation of the spin-wave dispersion from the assumed parabolicity used to determine $A$. In other words, it shows here that the exchange constant $A$ is obtained from an approximation of the Heisenberg model that assumes a small variation in the orientation of the local magnetic moments, which is only insufficiently fulfilled in many cases, especially for many $M_\mathrm{S}(T)$ measurements. Additionally, accurate modelling of $M_\mathrm{S}(T)$ data requires accounting for additional effects that emerge at the ultrathin limit, such as reduced dimensionality and confinement modes \cite{Erickson.1991}.  

% Outlook
\paragraph{Outlook}
In particular, we find that the analysis and interpretation of $M_\mathrm{S}(T)$ data for ultrathin films is far from trivial, as the underlying models contain numerous assumptions that are only partially satisfied \cite{Erickson.1991}. As a general consequence, when using $A$ to describe the exchange interaction in ultrathin films, one must be duly mindful of the implicit approximations, which depend on the relevant length scales or wave vectors. Therefore, when modelling ultrathin films, it may be advantageous to use $A$-values that are obtained by methods considering length scales similar to the problem at hand. In general, it is advisable to quantify the accuracy of a measurement of $A$ using a systematic assessment, i.e. values from methods with small wave vectors (e.g. BLS) should rather be considered as upper limit for $A$ while values from $M_\mathrm{S}(T)$ measurements can be considered as lower limit. For an improved determination of $A$ with the help of $M_\mathrm{S}(T)$ measurements, a numerical modelling of $M_\mathrm{S}(T)$ with the non-approximated spin wave dispersion including an accurate treatment of high wave vectors and dimensionality effects  could be beneficial, but this approach might easily lead to an over-parameterisation of the problem.  Additionally, our results would also improve the interpretation of other measured micromagnetic parameters such as DMI, whose extraction from experimental data typically includes the estimation of $A$.

\section*{Acknowledgments}
We acknowledge the support of the National Supercomputing Centre (NSCC), Singapore for computational resources. This work was supported by the SpOT-LITE program (A*STAR Grant No. A18A6b0057) funded by Singapore's RIE2020 initiatives, and by NUS funds (Grant No. A-0004544-00-00). Funding by the Deutsche Forschungsgemeinschaft within the CRC TRR173 \textit{Spin+X} (No.~268565370 (Projects B01 and B11)) and within the Priority Program SPP2137 \textit{Skyrmionics} (Project No.~403512431) is gratefully acknowledged.

\section{Appendix}%
\label{Appendix}%
%%%%%%%%%%%%%%%%%%%%%%%%%%%%%%%%%%%%%%%%%%%%%%%%%%%%%%%%%%%%
%%%%%%%%%%%%%%%%%%%% Ms(T) Measurements %%%%%%%%%%%%%%%%%%%%
%%%%%%%%%%%%%%%%%%%%%%%%%%%%%%%%%%%%%%%%%%%%%%%%%%%%%%%%%%%%

\subsection*{$\mathbf{M(T)}$ Measurements and Analysis}
\begin{figure}
	\centering
	\includegraphics[width=0.45\textwidth]{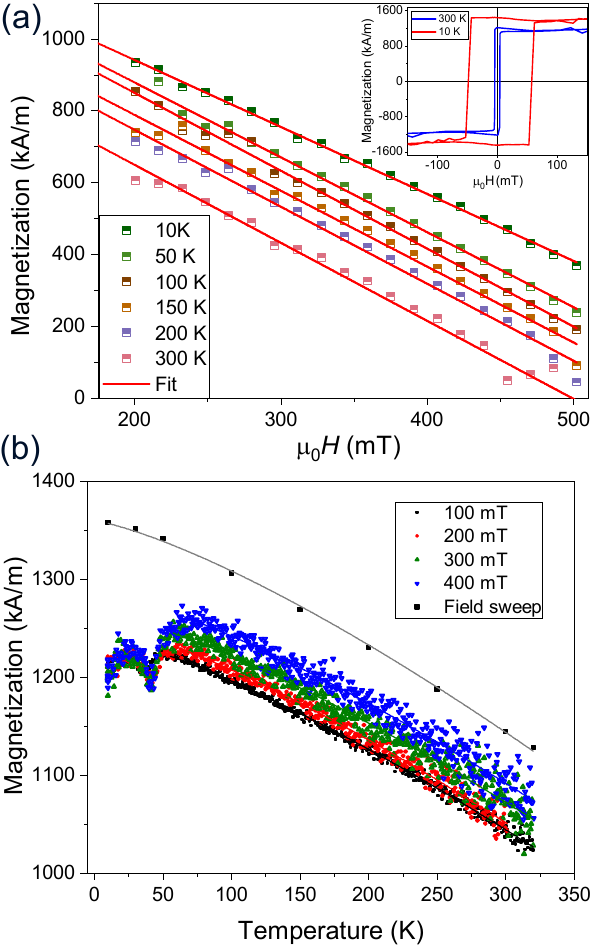}
	\caption{(a) M(H) measurements for Ir/Co(1)/Pt measured in OP geometry at fixed temperatures over 10$-$300 K for fields, $H>H_s$. The data were linearly fitted (red), and their extrapolation to ZF was used to determine $M_s$ for each temperature. Inset shows corresponding M(H) hysteresis loop measurements at 10 K (red) and 300 K (blue) (b) $M(T)$ measurements for Ir/Co(1)/Pt sample obtained in OP geometry at several fixed fields (100-400~mT), and from $M(H)$ extrapolation (black, from Fig.~\ref{M_T}). Deviations in $M(T)$ for $T \lesssim$ \SI{50}{K} are due to residual oxygen \cite{Gregory.1978}. Solid lines are fits to the 2D PSSW model, Eqn.~\eqref{m_reduction_pssw}.}
	\label{M_T_fields}
\end{figure}

\paragraph{$M(H)-T$ Method}
The $M_\mathrm{S}(T)$ data in Fig.~\ref{M_T} was obtained from $M(H)$ measurements over fields above saturation, $H>H_\mathrm{S}$, for several fixed temperatures. To remove substrate contributions, the measured $M(H)$ for $H>H_\mathrm{S}$ was fit to a straight line. The negative slope, resulting from the diamagnetic substrate, was removed by extrapolation to $H=0$. The magnetization was then calculated in intensive units (MA/m) by accounting for sample dimensions. The Fig.~\ref{M_T_fields}(a) shows the measured $M(H)$ data over fields above saturation at different fixed temperatures for the Ir/Co(1)/Pt sample in OP configuration.

\paragraph{$M-T$ Method: Fixed Field}
For comparative purposes, additional magnetometry measurements were performed to quantify $M(T)$ at fixed external fields. A key advantage of this fixed field method is that the spin-wave dispersion relation (i.e. the dipolar gap $\omega_0$) is unchanged during the measurement, which, in principle allows the use of Eq.~\ref{eq:2D} and Eq.~\ref{m_reduction_pssw} without any approximations. However, in this case, removing the substrate contribution to the measured $M(T)$ requires additionally a reference measurement of the bare substrate, whose dimensions and weight need to be identical to the sample of interest. Fig.~\ref{M_T_fields}(b) shows the measured $M(T)$ data at different fixed fields for the Ir/Co(1)/Pt sample in OP configuration. To circumvent the deviations observed for $T \lesssim 50$~K, known to arise from residual oxygen \cite{Gregory.1978}, $M(T)$ data for $T < \SI{70}{K}$ have been omitted from the fits (Fig.~\ref{M_T_fields}(b): solid lines). 

\paragraph{Dipolar Gap}
As the thermal occupation of states is highest for the fundamental spin-wave mode, the dipolar gap $\omega_0$ is a crucial parameter for the reliable use of $M_\mathrm{S}(T)$ models for ultrathin films (Eqs.~\eqref{eq:2D}, \eqref{m_reduction_pssw}). 
First, for $M(H)$ measurements involving the extrapolation method, $\omega_0$ was estimated using the centre-point of the covered field range as $\upmu_0 H_\mathrm{ext}$, together with the $M_\mathrm{S}$ value obtained from the $M_\mathrm{S}(T)$ extrapolation, and the anisotropy field $\upmu_0 H_\mathrm{U}$ extracted from BLS measurements. Using these, we have \cite{Kittel.1948}
\begin{equation}
	\label{omega0}
	\omega_0 = \gamma \upmu_0 (H_\mathrm{ext} + H_\mathrm{U} - M_\mathrm{S})\quad,
\end{equation}
with $\upmu_0 H_\mathrm{ext} = \SI{300}{mT}$. Meanwhile, for $M(T)$ measurements at fixed fields, $\omega_0$ can be reliably determined given the constancy of $\upmu_0 H_\mathrm{ext}$. 

\paragraph{PSSW Fit Method}
To estimate $A$ using the $M_\mathrm{S}(T)$ PSSW model, we perform the following iterative procedure to account for the implicit $A$-dependence of the PSSW modes $\omega_1$ and $\omega_2$. We start with the above mentioned estimation of the fundamental frequency $\omega_0$, and use it to evaluate the strict 2D model (Eqn.~\eqref{eq:2D}). This gives a lower estimate of $A$, which we then use, together with the $M_\mathrm{S}(T=0)$ estimate and the film thickness, $t$, to calculate $\omega_1$ and $\omega_2$. These frequencies are used to fit the more precise 2D PSSW model (Eqn.~\eqref{m_reduction_pssw}) up to $n=2$ to the data. We iterate this fit twice using recalculated PSSW frequencies. The resulting $A$ varied by $<$\SI{1}{\%}  between the second and third iterations.

\paragraph{$A$-Values Comparison}
The calculated values of $\omega_0$ are listed in Tbl.~\ref{tab:M_T} for the different $\upmu_0 H_\mathrm{ext}$ values  used in $M(T)$ and $M(H)$ measurements, together with the resulting $A$ values obtained from fits to the Bloch's $T^{3/2}$ law, and the 2D PSSW model (Eqn.~\eqref{m_reduction_pssw}). The $A$-values are in good agreement across different $\upmu_0 H_\mathrm{ext}$ values, confirming that both $M(H)$ and $M(T)$ magnetometry data can be used to determine $A$ equally well, with appropriate estimation of $\omega_0$. 

\begin{table}
	\renewcommand{\arraystretch}{2}
	\begin{ruledtabular}
		\begin{tabular}{c|ccc}
			$\upmu_0 H_{\rm ext}$ & \parbox[c][3em][c]{1.5cm}{$\omega_0/(2\pi)$\\ $(\SI{}{GHz})$}  & \parbox[c][3em][c]{1.5cm}{$A_{T^{3/2}}$\\ $(\SI{}{pJ/m})$} & \parbox[c][3em][c]{1.5cm}{$A_\mathrm{PSSW}$ \\$\SI{}{(pJ/m})$}\\  
			\hline 
			$\SI{100}{mT}$& $\SI{20}{}$ & $\SI{4.41 \pm 0.50}{}$  & $\SI{6.98 \pm 0.50}{}$   \\
			\hline
			$\SI{200}{mT}$ & $\SI{23}{}$ & $\SI{4.35 \pm 0.50}{}$ & $\SI{6.79 \pm 0.50}{}$ \\
			\hline
			$\SI{300}{mT}$ & $\SI{25}{}$ & $\SI{4.56 \pm 0.50}{}$ & $\SI{7.06\pm 0.50}{}$ \\
			\hline
			$\SI{400}{mT}$ & $\SI{28}{}$ & $\SI{4.61 \pm 0.50}{}$ & $\SI{7.03 \pm 0.50}{}$  \\
			\hline
			Field Sweep & $\SI{25}{}$ & $\SI{4.51 \pm 0.50}{}$ & $\SI{7.05 \pm 0.50}{}$ \\
		\end{tabular}
	\end{ruledtabular}
	\caption{\label{tab:M_T}
		Magnitude of dipolar gap $\omega_0$ (Eqn.~\ref{omega0}) for the $M(T)$ and $M(H)$ measurements performed on sample Ir/Co(1)/Pt over varying external fields in OP configuration, and the resulting $A$ values obtained from the 3D Bloch law and 2D PSSW $M_\mathrm{S}(T)$ models.}
\end{table}
\renewcommand{\arraystretch}{1}

%%%%%%%%%%%%%%%%%%%%%%%%%%%%%%%%%%%%%%%%%%%%%%%%%%%%%%%%%%%%
%%%%%%%%%%%%%%%%%%%% Exchange in Bilayers %%%%%%%%%%%%%%%%%%%%
%%%%%%%%%%%%%%%%%%%%%%%%%%%%%%%%%%%%%%%%%%%%%%%%%%%%%%%%%%%%
\subsection*{Numerical modeling approach: Effective Exchange Stiffness for FM Bilayers}%
\begin{figure}
	\centering
	\includegraphics[width=0.48\textwidth]{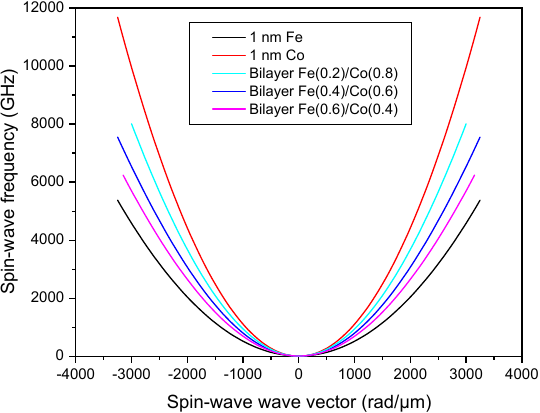}
	\caption{Spin-wave dispersion of pure Co, pure Fe, and Fe/Co bilayer films, obtained by numerical modeling (parameters in text). Parabolic fits to the data are used to assess the validity of several analytical models to estimate $A$ for bilayers.}
	\label{bilayer_dispersion}
\end{figure}
\paragraph{Bilayer Model Motivation}
One of the five samples used in this work is a bilayer FM, Fe/Co. To model spin-wave dispersions of thin films with inhomogeneous material parameters across their thickness, numerical calculations \cite{Hillebrands.1990} and numerical modeling \cite{Kostylev.2014} may serve as valuable tools. First, we note that for the other four single magnetic layer films, modelling using ref.~\cite{Kostylev.2014} gives $A$-values consistent with Eqn.~\eqref{Eq:Dispersion_sym}, within errors bars. For example, using the ref.~\cite{Kostylev.2014} approach to localize the uniaxial anisotropy only on one surface, gives similar results to the effective volume anisotropy approach (Eqn.~\eqref{Eq:Dispersion_sym}), as the exchange interaction enforces homogeneous dynamics across the ultra-low film thickness. 

\paragraph{Bilayer Model Methods and Setup}
To model the effective dispersion of the Fe/Co bilayer and determine its effective exchange parameter $A$, we compare results from Eqn.~\eqref{Eq:Dispersion_sym} to an advanced dispersion model, with arbitrary distribution of $M_\mathrm{S}$, $H_\mathrm{U}$, and $A$ over the film thickness \cite{Kostylev.2014}. We thereby verify that the spin-wave dispersion in ultrathin FM bilayers can be well-described using an effective $A$.
To do so, first, we numerically calculated the dispersion of the fundamental mode for bilayer Fe($t_\mathrm{Fe}$)/Co($t-t_\mathrm{Fe}$) with $t=\SI{1}{nm}$, over IP wave vectors \SI{-3000}{rad \per \upmu m} to $+\SI{3000}{rad \per \upmu m}$, using expected values for Co: $M^\mathrm{(Co)}_\mathrm{S} = \SI{1050}{kA \per m}$, $A_\mathrm{Co} = \SI{20}{pJ \per m}$, and for Fe, $M^\mathrm{(Fe)}_\mathrm{S} = \SI{1600}{kA \per m}$, and $A_\mathrm{Fe} = \SI{14}{pJ \per m}$. We fit the data to a function of the form 
\begin{equation}
	\label{bilayer_disp}
	f_\mathrm{SW}(k)=\frac{\gamma}{2 \pi} \frac{2A}{M}\cdot k^2=\beta_\mathrm{num} k^2. 
\end{equation}
Here, $\beta_\mathrm{num}$ is a proportionality factor given by $\beta_\mathrm{num} = (\gamma \upmu_0 /2 \pi)\cdot \lambda_\mathrm{ex,eff}$, where $\lambda_\mathrm{ex,eff}$ is the effective spin-wave stiffness. 
In order to compare different analytical approaches to the numerical data, we evaluate the following model relations between the individual and bilayer exchange stiffnesses:
\begin{equation}
	\label{beta1}
	\beta_1 = \beta_\mathrm{Fe} \cdot \frac{t_\mathrm{Fe}}{t} + \beta_\mathrm{Co} \cdot \frac{t_\mathrm{Co}}{t}
\end{equation}
\begin{equation}
	\label{beta2}
	\beta_2 = \frac{\gamma}{\pi} \frac{(A_\mathrm{Fe}+A_\mathrm{Co})/2}{(M_\mathrm{S}^\mathrm{(Fe)} \cdot t_\mathrm{Fe}+M_\mathrm{S}^\mathrm{(Co)} \cdot t_\mathrm{Co})/t}
\end{equation}
\begin{equation}
	\label{beta3}
	\beta_3 = \frac{\gamma}{\pi} \frac{A_\mathrm{Fe}\cdot t_\mathrm{Fe} + A_\mathrm{Co} \cdot t_\mathrm{Co}}{M_\mathrm{S}^\mathrm{(Fe)} \cdot t_\mathrm{Fe}+M_\mathrm{S}^\mathrm{(Co)} \cdot t_\mathrm{Co}}.
\end{equation}
Here, $\beta_\mathrm{Fe}$ and $\beta_\mathrm{Co}$ are the curvatures fitted to the numerically evaluated dispersion data for \SI{1}{nm} thick Fe and Co films, respectively. 

\paragraph{Bilayer Model Results}
Tbl.~\ref{tab:exchange} lists the values obtained for the different effective exchange models. It is apparent that the $\beta_3$ model (Eqn.~\eqref{beta3}) best describes the numerically obtained exchange stiffness for bilayer films. This result also underlines that within the micromagnetic framework, the parabolic dispersion approximation for $M(T)$ analysis can be extended to bilayer samples, while appropriately accounting for the thickness-weighting of $A$ and $M_\mathrm{S}$, as described by $\beta_3$ (Eqn. \ref{beta3}). We have also verified that the strength of exchange coupling between the two layers plays at best a minor role. As long as the layers were FM coupled, we observed no effect of the interlayer coupling strength on the effective dispersion, in line with our T-dependent texture evolution studies on similar samples \cite{Chen.2022}.
%As long as the layers were FM-coupled, we observed no effect of the interlayer coupling strength on the effective dispersion. 

\begin{table}
	\renewcommand{\arraystretch}{1.5}
	\begin{ruledtabular}
		\begin{tabular}{c|cccc}
			$t_\mathrm{Fe}$ & $\beta_\mathrm{num}$ & $\beta_1$ & $\beta_2$ & $\beta_3$\\  
			(nm) & \multicolumn{4}{c}{($10^{-4}$ \SI{}{GHz \per (rad \per \upmu m)})}  \\ 
			\hline 
			0.2&9.03&9.85&8.21&9.08\\
			0.4&7.36&8.66&7.50&7.76\\
			0.6&6.42&7.47&6.90&6.66\\
		\end{tabular}
	\end{ruledtabular}
	\renewcommand{\arraystretch}{1}
	\caption{\label{tab:exchange}
		Effective exchange stiffness values for 1~nm Fe/Co bilayers with varying compositions ($t_{\rm{Co}} = 1 - t_{\rm{Fe}}$), obtained by fitting the spin-wave dispersion using analytical models $\beta_{1,2,3}$ (Eqns.~\eqref{beta1}-\eqref{beta3}). Comparison to the numerical model, $\beta_\mathrm{num}$, yields best agreement for $\beta_3$ (Eqn.~\ref{beta3}).}
\end{table}

%%%%%%%%%%%%%%%%%%%%%%%%%%%%%%%%%%%%%%%%%%%%%%%%%%%%%%%%%%%%
%%%%%%%%%%% Dispersion Models: Full v. Parabolic %%%%%%%%%%%
%%%%%%%%%%%%%%%%%%%%%%%%%%%%%%%%%%%%%%%%%%%%%%%%%%%%%%%%%%%%

\subsection*{Full and Parabolic Spin-Wave Dispersion Models}%

\paragraph{Full/Parabolic SW Model Intro}
Since each thermally excited magnon reduces $M_\mathrm{S}(T)$ by $g \upmu_\mathrm{B}$, the resulting $M_\mathrm{S}(T)$ is given as
\begin{equation}
	\label{m_reduction_general}
	\begin {split}
	M_\mathrm{S}(T) = M_\mathrm{S}(0)-g \upmu_\mathrm{B} \sum_{\vec{k}} n(\vec{k},T) \\ = M_\mathrm{S}(0)-g \mu_\mathrm{B}\int_0^\infty N(\omega,T) D(\omega) \mathrm{d}\omega.
\end{split}
\end{equation}
where $n(\vec{k},T)$ is the magnon density per unit volume, which, can be expressed as the product of the DOS $D(\omega)$ and the Bose-Einstein distribution factor $N(\omega)$ (Eqn.~\eqref{thermal_population}).
This is the basis of the model leading to Eqs.~\eqref{m_reduction_bloch} and \eqref{m_reduction_pssw}. 
Meanwhile, the full spin-wave dispersion for the Heisenberg model on a 2D square lattice can be described as \cite{Bible.2017}
\begin{equation}
\label{dispersion_full}
\omega(k)_\mathrm{full} =  \omega_0 + \frac{2 \gamma \upmu_0 \lambda_\mathrm{ex}}{a^2} \left( 2 - \cos(k_x a) - \cos(k_y a) \right)
\end{equation}
where $a$ is the atomic lattice constant, and the 2D lattice is oriented in the $xy$-plane with spin wave wave vector components $k_x$ and $k_y$. The ensuing parabolic approximation to the dispersion relation is then given by
\begin{equation}
\label{dispersion_parabolic}
\omega(k)_\mathrm{parabolic} =  \omega_0 + \frac{\gamma \upmu_0 \lambda_\mathrm{ex}}{a^2} \left(k_x^2 + k_y^2 \right) a^2 \,.
\end{equation}
\begin{figure}
\centering
\includegraphics[width=0.48\textwidth]{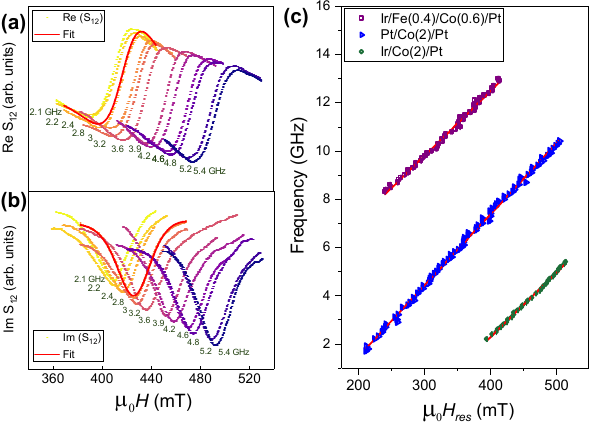}
\caption{(a-b) Exemplary real and imaginary parts of complex transmission parameter, $S_\mathrm{12}$ for sample Ir/Co(2)/Pt, measured by VNA-FMR in OP sample geometry at fields above saturation, $H > H_{\rm{S}}$. (c) Dispersion of resonance field, $H_{\rm{res}}$ with frequency, determined from Lorentzian fits to resonance spectra (e.g. (a,b)) for samples Ir/Fe(0.4)/Co(0.6)/Pt (purple), Pt/Co(2)/Pt (blue) and Ir/Co(2)/Pt (green). Overlaid lines are linear fits to the Kittel formula (Eqn.~\ref{eq:Kittel}).}
\label{FMR}
\end{figure}
\paragraph{Full/Parabolic SW Model Comparison}
The dispersions of the full model and the parabolic approximation are compared in Fig.~\ref{dispersion-deviation}a), for a representative set of parameters (see main text). The increasing deviation between the models at larger frequencies hints at the complexity in accurately describing the measured $M_\mathrm{S}(T)$ dependence at room temperature. 
This can be further understood by evaluating the corresponding DOS, $\rho(\omega)$, performed here numerically. The DOS is given by the number of available states per frequency interval $\mathrm{d}\omega$ around $\omega$ such that
\begin{equation}
\label{DOS}
\rho(\omega) = \frac{1}{S} \sum_{\vec{k} \in K} \delta(\omega-\omega(\vec{k}))
\end{equation}
where $S$ and $K$ are the total surface area in real and $k$-space, respectively. It is evident that the full Heisenberg model (Eqn.~\eqref{dispersion_full}) gives a larger DOS compared to the parabolic approximation (Eqn.~\eqref{dispersion_parabolic}).

\paragraph{SW Model with DMI}
To calculate the DOS in the presence of DMI, a $k$-linear term is added to the dispersion relation \cite{Di.2015b,Nembach.2015,Belmeguenai.2015}
\begin{equation}
\omega_\mathrm{DMI}(k) = \frac{2 \gamma}{M_\mathrm{S}} D k_x \,.
\end{equation}
Evidently, the DMI-induced dispersion modification does not significantly influence the DOS (Fig.~\ref{dispersion-deviation}b, blue curve). Thus, the $M_\mathrm{S}(T)$ dependence is predicted to be very similar for magnetometry measurements in IP (interfacial DMI contribution is expected) and OP (interfacial DMI influence can be neglected) field orientations which is in very good agreement with our experimental findings.
%%%%%%%%%%%%%%%%%%%%%%%%%%%%%%%%%%%%%%%%%%%%%%%%%%%%%%%%%%%%
%%%%%%%%%%%%%%%%%%%% FMR Measurements %%%%%%%%%%%%%%%%%%%%
%%%%%%%%%%%%%%%%%%%%%%%%%%%%%%%%%%%%%%%%%%%%%%%%%%%%%%%%%%%%

\subsection*{Ferromagnetic Resonance Measurements}
\paragraph{FMR Methods}
Ferromagnetic resonance (FMR) measurements were performed to determine the gyromagnetic ratio, $\gamma$, using a home-built broadband vector network analyzer (VNA) magnetic absorption spectroscopy  setup. The samples were mounted on a coplanar waveguide (CPW) in OP field geometry, and the complex microwave transmission parameter S$_{12}$ was recorded at fixed microwave frequencies, $f$ over 1-14  GHz, as a function of the OP magnetic field ($\mu_0$H$_{\mathrm{ext}}$, up to $\sim$ 0.55 T). The resulting real and imaginary parts of S$_{12}$, exemplified in Fig.~\ref{FMR}(a-b) for sample Ir/Co(2)/Pt, were fit to the Lorentzian lineshape functions to determine the resonance field, $\mu$$_{0}$$H$$_{res}$, for each frequency.
\paragraph{FMR Results}
The resulting $H_{res}$-$f$ dispersion plot, shown in Fig.~\ref{FMR}(c) for the three measured samples, can be fit by the Kittel formula \citep{Kittel.1948} for OP geometry,
\begin{equation}
\label{eq:Kittel}
f  =  \frac{\mu_0 \gamma}{2\pi}(H_\mathrm{res}-M_\mathrm{eff})\;.
\end{equation}
Here, $M_\mathrm{eff}=M_\mathrm{S}-H_\mathrm{U}$, the effective magnetization, includes contributions from the uniaxial anisotropy field $H_\mathrm{U}$. The three $H_{\rm{res}}-f$ plots in Fig.~\ref{FMR}(c) show a large variation in their $y$-intercepts, which reflects the anisotropy evolution across samples. Meanwhile, their slopes, proportional to the quantity of interest, $\gamma$, exhibit marginal ($\sim \pm 5\%$) variation across samples. The observed magnitude of variation of $\gamma$  across samples, expected to arise from variations in the $g$-factor, is consistent with previous reports on multilayer films \citep{Beaujour.2007}. The measured $\gamma$ values are summarized in Tbl.~\ref{tab:summary}, and used in determining $A$ for both BLS and $M(T)$-based methods.
%
%\section{Additional comments}
%\input{Exch_X_Comments}%
%
%\bibliographystyle{apsrev4-1}
\end{document}